\RequirePackage{fix-cm}
\documentclass[smallcondensed]{svjour3}     
\smartqed  
\usepackage{graphicx}
\usepackage[compat=1.1.0]{tikz-feynman}
\usepackage{amsmath}
\usepackage{placeins}
\usepackage{authblk}

\begin{document}

\title{Design of a Multiple-Reflection Time-of-Flight Mass Spectrometer for Barium-tagging}

\author[1]{K. Murray}
\author[2,3]{J. Dilling}
\author[2,4]{R. Gornea}
\author[1]{Y. Ito\footnote{Now at JAEA}}
\author[4]{T. Koffas}
\author[2]{A.A. Kwiatkowski}
\author[2,3]{Y. Lan}
\author[2]{M.P. Reiter}
\author[5,6]{V. Varentsov}
\author[1,2]{T. Brunner}
\author[ ]{with the nEXO collaboration}


\authorrunning{K. Murray} 

\institute{Kevin Murray \\\email{kevin.murray2@mail.mcgill.ca}\\ The nEXO Collaboration \\\email{thomas.brunner@physics.mcgill.ca}}

\affil[1]{McGill University, Montreal, Canada}
\affil[2]{TRIUMF, Vancouver, Canada}
\affil[3]{University of British Columbia, Vancouver, Canada}
\affil[4]{Carelton University, Ottawa, Canada}
\affil[5]{Facility for Antiproton and Ion Research in Europe (FAIR GmbH), Darmstadt, Germany}
\affil[6]{Institute for Theoretical and Experimental Physics, Moscow, Russia}
%

\date{August 2019}

\titlerunning{Design of an MR TOF for Barium-tagging}

\maketitle


\begin{abstract}
The search for neutrinoless double beta decay requires increasingly advanced methods of background reduction. A bold approach to solving this problem, in experiments using $^{136}$Xe, is to extract and identify the daughter $^{136}$Ba ion produced by double beta decay. Tagging events in this manner allows for a virtually background-free verification of double beta decay signals. Various approaches are being pursued by the nEXO collaboration to achieve Ba-tagging. A Multi-Reflection Time-of-Flight Mass Spectrometer (MR TOF) has been designed and optimized as one of the ion-identification methods, where it will investigate the ion-extraction efficiency, as well as provide further identification of the Ba isotope. The envisioned mode of operation allows the MR TOF to achieve a quickly adjustable mass-range and resolution, with simulations suggesting that a mass-resolving power of 140,000 is within reach. This work will discuss the MR TOF design and the methods employed to simulate and optimize it.
\end{abstract}

\section{Introduction}
\label{intro}
\begin{figure}
    \centering
    \includegraphics[width = \textwidth]{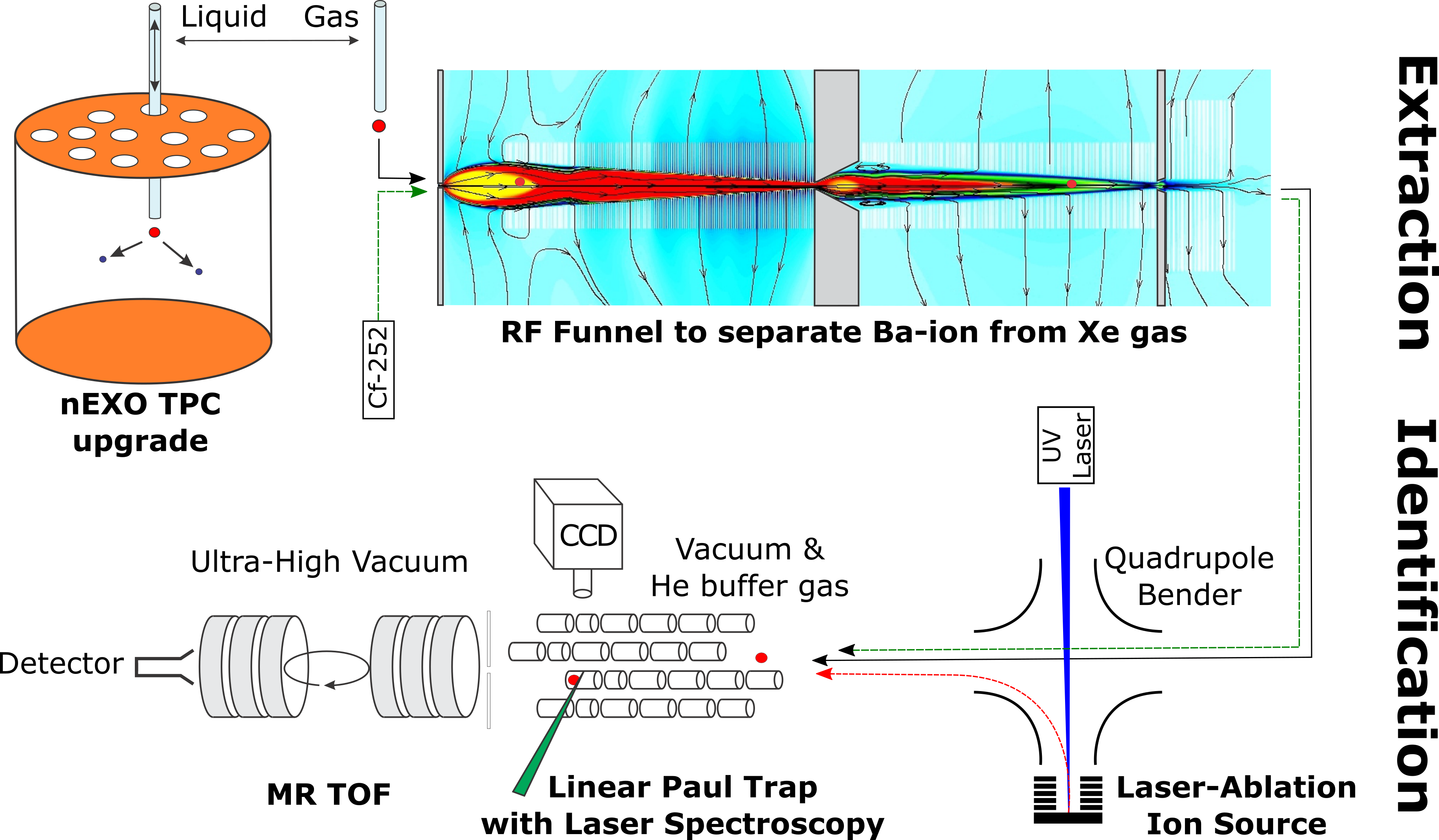}
    \caption{The Ba-tagging setup envisioned as a potential upgrade to the nEXO experiment. Ions are extracted from the TPC with a phase changing capillary. They are then separated from the neutral Xe gas with a system of three deferentially pumped RF funnels, for which the design, combined with the results of gas dynamic simulations, are shown in the upper right corner. They are then cooled and bunched with an LPT, where the Ba ion is identified with laser spectroscopy, and ejected into an MR TOF for systematic studies of the ion-extraction technique. Radioactive (green line) and laser-ablation (red line) ion sources will be used to test the offline setup.}
    \label{fig:batag}
\end{figure}

The Multi-Reflection Time-of-Flight Mass Spectrometer (MR TOF), as first proposed by H. Wollnik et. al. \cite{WOLLNIK1990267}, is becoming increasingly popular at radioactive ion beam facilities across the world \cite{hirsh2016first,jesch2017mr,chauveau2016pilgrim,schury2017first,wienholtz2013masses,PhysRevLett.120.152501,leistenschneider2018dawning}. This is in part due to its excellent performance as an isobar separator, as well as its ability to achieve a mass-resolving power in excess of $10^5$, with only milliseconds of measurement time \cite{wolf2013isoltrap}. Moreover, an MR TOF is capable of broadband mass measurements with high transmission efficiency ($>50\%$), and single ion sensitivity \cite{dickel2015high}. These qualities make the device suitable for use in an advanced method of background suppression, currently under development, called Ba-tagging. $^{136}$Xe double beta decays ($\beta \beta$) to $^{136}$Ba, which is a stable isotope. This presents the opportunity to extract the daughter Ba ion from the detector volume, and identify it to corroborate detector signals \cite{mcdonald2018demonstration,mong2015spectroscopy,twelker2014apparatus,chambers2018imaging}. For a neutrinoless double beta decay experiment ($0\nu\beta\beta$), such as nEXO \cite{kharusi2018nexo}, a future upgrade to take advantage of such a technique would eliminate all backgrounds, except for those created by ordinary $\beta\beta$, and greatly increase the experimental sensitivity to $0 \nu \beta \beta$  \cite{albert2018sensitivity}. A candidate $0 \nu \beta \beta$ event, in the nEXO liquid Time-Projection Chamber (TPC), can be localised to a volume with a characteristic dimension of a few mm \cite{albert2014improved}, with an ion fraction $>76\%$ \cite{albert2015measurements}. In a potential Ba-tagging scenario, the volume surrounding the decay daughter is extracted with a capillary in which the phase changes from liquid to gas. The ions extracted from the TPC are then separated from neutral Xe gas with a radio frequency ion-funnel (RF funnel) \cite{brunner2015rf}. They are then cooled and bunched with a Linear Paul Trap (LPT), in which the Ba ion is identified with laser spectroscopy. Ion bunches are ejected from the LPT into an MR TOF for further analysis. An overview of the Ba-tagging setup is shown in Fig. \ref{fig:batag}. The ion-extraction scheme, using the RF funnel, LPT and MR TOF will first be tested with ``offline" radioactive and laser-ablation ion sources. For further discussion on the testing setup, see Refs. \cite{brunner2017searching,VVref1,VVref2,VVref3}. Once Ba-extraction from Xe gas has been demonstrated, the scheme will be applied to a future upgrade of nEXO. \newline

The MR TOF described in this work, will be used primarily for broadband mass measurements of the ion-extraction process. However, it also has the potential to provide secondary identification of the Ba ion, if sufficient mass-resolving power and efficiency is achieved ($m/\Delta m>100,000$).  

\section{The Multi-Reflection Time-of-Flight Mass-Spectrometer}
\label{sec:2}
Even though the working principle of an MR TOF is quite simple, there are multiple ion-optical considerations that make its optimization and operation challenging. The following will outline the machinery used to calculate and manipulate the mass-resolving power. \newline

Ions with a mass-to-charge ratio $m/q$, that are accelerated by a potential energy gradient $\Delta U$, are given a velocity of 
\begin{equation}
v = \sqrt{\frac{q}{m}2\Delta U}.
\end{equation}
In principle, the time taken $t$, to traverse a path of length $L$, is proportional to the square root of the mass-to-charge ratio, $t = L/v \propto \sqrt{m/q}$. Hence, the mass-resolving power can then be expressed as $R \equiv m/ \Delta m = t/(2\Delta t)$. In reality, an ion bunch has a finite energy and spatial distribution when accelerated by $\Delta U$. This results in a couple of effects that increase $\Delta t$ and reduce the resolving power. Firstly, inside the LPT some ions must reverse their momentum before they can accelerate into the MR TOF; this results in an effect called the turnaround time \cite{chandezon1994new}. Secondly, ions are only given approximately the same kinetic energy, since they start at slightly different heights on the potential energy slope. Lastly, ions may be displaced from the optical axis, which contributes to differences in path length. \newline

\begin{figure}
    \centering
    \includegraphics[width = \textwidth]{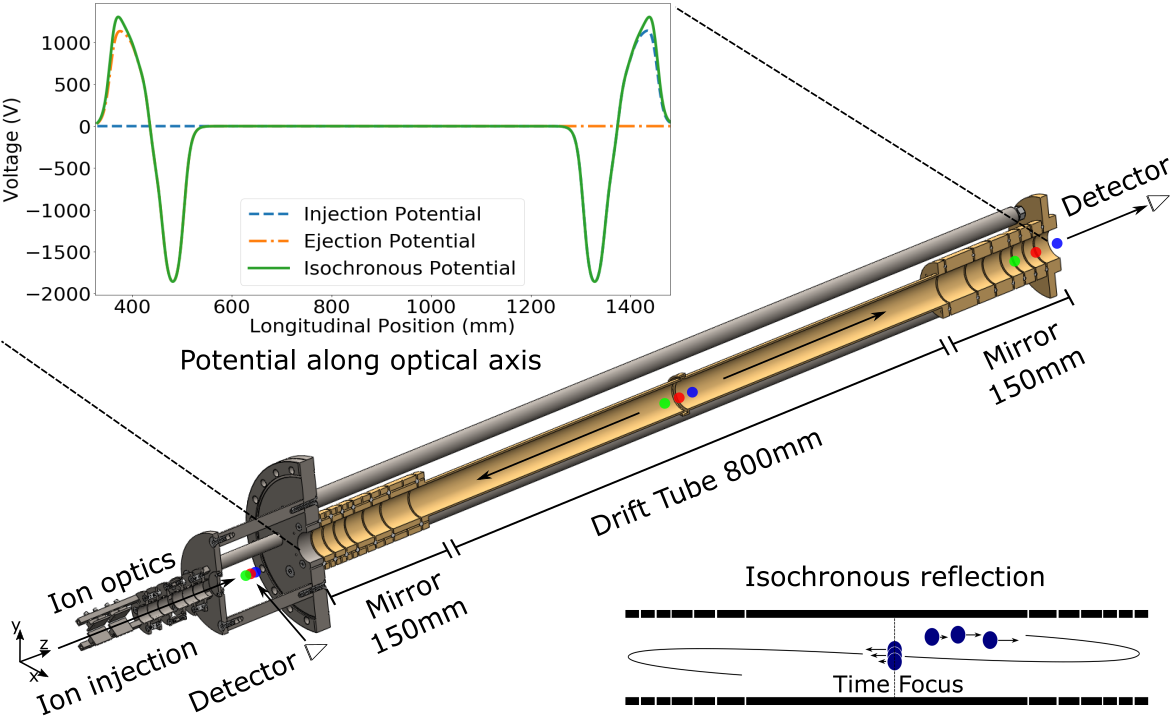}
    \caption{The MR TOF consists of 2 electrostatic mirrors of 6 electrodes each, and a central drift tube. Ion bunches from the LPT pass through 2 sets of $x-y$ steerers, to correct for misalignment, and 2 einzel lenses, to control the beam focus. A small collimator plate placed after the second einzel lens limits the maximum beam size. The MR TOF is operated in a turn independent mode, by shifting the time focus to and from the centre of the drift tube with the first and last reflections, and making all intermediate reflections isochronous.}
    \label{fig:mrtof_main}
\end{figure}

In an MR TOF, ion bunches are injected between 2 electrostatic mirrors for $N$ turns, so that the flight path can be extended within a compact space \cite{WOLLNIK1990267}. The mirrors consist of 6 electrodes each and are located on both ends of a central drift tube. This configuration is based on the ISOLDE design \cite{wolf2013isoltrap}, and the components will be machined from aluminium and subsequently gold-plated. Two sets of $x-y$ steerers and einzel lenses will be placed between the LPT and MR TOF to correct for any misalignment and control the beam focus, see Fig. \ref{fig:mrtof_main}. The resolving power can be modeled with \cite{wolf2013isoltrap}
\begin{equation}
R = \frac{t_0 + NT}{2\sqrt{(\Delta t_0)^2+(N\Delta T)^2}},
\label{resolvingpower}
\end{equation}
where $t_0$ is the TOF from the LPT to the detector, without any turns in the MR TOF. $T$ is the TOF for a single turn, $\Delta t_0$ is the turn-around time and $\Delta T$ is the contribution to TOF dispersion per turn. \newline

The ion transport in this work is modeled using transfer matrix formalism, which is used to describe the effects of ion optical devices and drift regions on an ion's trajectory \cite{wollnik2012optics}. The properties of an ion's trajectory can be described at any point in its flight with the vector $X = [x,a,y,b,\delta,t']$, where $x$ and $y$ are transverse displacements, $a=(\partial x/\partial t)/(\partial z/\partial t)$ and $b=(\partial y/\partial t)/(\partial z/\partial t)$ are transverse angles, all of which are relative to a reference trajectory along the $z$ axis \cite{yavor2009optics}, $\delta = (K-K_r)/K_r$ is the deviation of an ion's kinetic energy $K$ from the reference ion's kinetic energy $K_r$. Likewise, $t' = t-t_r$ is the difference in TOF of the considered ion from the reference ion. After a reflection in the MR TOF, ion properties can be expressed as \cite{dickel2015high}

\begin{align}
\label{MRTOFREF1}
x =& (x|x)x_0+(x|a)a_0+(x|x\delta)x_0\delta+(x|a\delta)a_0\delta+...\\
a =& (a|x)x_0+(a|a)a_0+(a|x\delta)x_0\delta+(a|a\delta)a_0\delta+...\\
y =& (y|y)y_0+(y|b)b_0+(y|y\delta)y_0\delta+(y|b\delta)b_0\delta+...\\
b =& (b|y)y_0+(b|b)b_0+(b|y\delta)y_0\delta+(b|b\delta)b_0\delta+...\\
t'=& (t|\delta)\delta+(t|xx)x_0^2+(t|x a)x_0 a_0+ (t|a a)a_0^2+(t|yy)y_0^2+(t|yb)y_0b_0\nonumber\\
  +& (t|bb)b_0^2 + (t|\delta\delta)\delta^2+(t|\delta\delta\delta)\delta^3+(t|\delta\delta\delta\delta)\delta^4+(t|xx\delta)x_0^2\delta\nonumber\\
  +&(t|xa\delta)x_0a_0\delta+(t|aa\delta)a_0^2\delta+ (t|yy\delta)y_0^2\delta+ (t|yb\delta)y_0b_0\delta\nonumber\\
  +&(t|bb\delta)b_0^2\delta+...,
\label{MRTOFREF2}
\end{align}

where the measurement point is taken to be the midplane between the coaxial mirrors. Optimization of the MR TOF can be achieved through precise manipulation of the coefficients in eqs. (\ref{MRTOFREF1}-\ref{MRTOFREF2}). For example, $(x|x)$ controls the parallel-to-point behaviour of the ion bunch through a reflection. If $(x|x)=0$, the final $x$ does not depend on initial $x$, i.e., the ions are focused to a point in the $x$ plane. Moreover, the MR TOF can correct for the spread in kinetic energy through $(t|\delta)$, $(t|\delta \delta)$, $(t|\delta \delta \delta)$ etc. However, the manner in which these coefficients will be manipulated depends on the MR TOF's operation mode.

\section{Operation Mode and Mass Range}
\label{sec:3}

It is clear that regardless of individual parameters, ions with the same $m/q$ must arrive at the detector with as little dispersion in time as possible. This concept is known as a time-focus. The LPT creates a time focus that is a few cm from the exit of the trap. In typical MR TOF operation, the time-focus is shifted onto the detector over the desired number of turns $N$. This method requires a re-tuning of the mirror potentials, or adjustment of the ion's kinetic energy in the case of the ISOLDE design, should $N$ change.\newline

The mass range is inversely proportional to the resolving power, since heavier slower ions may begin to be overlapped by lighter quicker ones, to the extent that peaks in the final spectrum may not correspond to the same $N$. This is an important consideration for the MR TOF as a broadband mass analyzer. The unambiguous mass range can be expressed as \cite{dickel2017al}

\begin{equation}
\frac{(m/q)_{\text{max}}}{(m/q)_{\text{min}}} = \left(\frac{N+\lambda_{\text{inj}}}{N+\lambda_{\text{inj}}-(1-\lambda_{\text{mir}})}\right)^2,
\label{eq:mass_acc}
\end{equation}
where $t_{inj} = \lambda_{\text{inj}} T$ is the TOF from the LPT to the beginning of the exit mirror, and $t_{\text{mir}} = \lambda_{\text{mir}} T$ is the TOF spent in the mirror for a single reflection. The unambiguous mass range depends strongly on $N$. Thus, it is preferable that the mirror potentials do not need to be re-tuned when the turn number is changed, so that the mass range and resolving power are easier to adjust. This is achieved through the method of time-focus shifting, as discussed in \cite{dickel2017al}, in which all intermediate reflections are tuned to be isochronous, whilst the first and last reflections are tuned to shift the time-focus to and from the centre of the drift tube, as shown in Fig. \ref{fig:mrtof_main}.\newline

\section{Simulation and Optimization}
\label{sec:3}
\begin{figure}
    \centering
    \includegraphics[width = \textwidth]{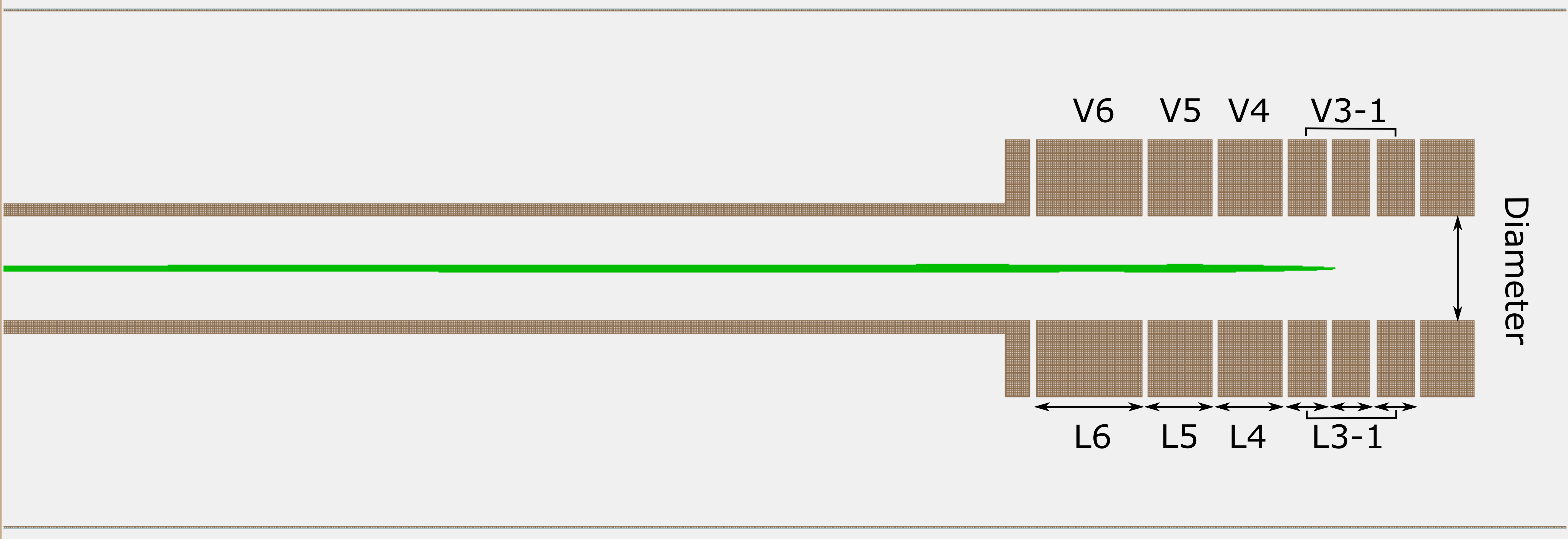}
    \caption{SIMION simulation for a single reflection in the MR TOF using simulated ions from a LPT. A reflection in the MR TOF is used to calculate the relevant coefficients in eqs. (\ref{MRTOFREF1}-\ref{MRTOFREF2}), to optimize the mirror potentials and geometry. The parameters V1-6 refer to the applied voltages, and L1-6 refer to electrode length.}
    \label{fig:reflection}
\end{figure}
The MR TOF has been optimized and simulated and with SIMION 8.1 \cite{DAHL20003}, using simulated ions from an LPT currently under design at TRIUMF \cite{instance1290}. The simulated ion bunch used has a mean kinetic energy of 1163 eV with a Full Width Half Maximum (FWHM) of 23 eV. There are 6 potentials in the MR TOF mirrors that must be optimized in order to minimize 6 coefficients in eqs. (\ref{MRTOFREF1}-\ref{MRTOFREF2}). Moreover, the MR TOF has several physical parameters that may be adjusted, namely the electrode length, spacing and radius, which is referred to collectively as a geometry, see Fig. \ref{fig:reflection}. The procedure for optimization is to select multiple geometries of interest and optimize the mirror potentials, so that their performance may be compared. Due to the large computing requirements, this was performed on the Compute Canada High Performance Computing clusters \cite{1742-6596-341-1-012001}. To optimize the mirror potentials, SIMION's native Nelder-Mead search algorithm was used. The algorithm takes in a scalar metric to minimize $\chi$, which is calculated as a weighted sum of the coefficients to minimize
\begin{equation}
\chi = \omega_1 (x|x) + \omega_2 (y|y) + \omega_3 (t|\delta) + \omega_4 (t|\delta \delta) + \omega_5 (t|\delta \delta \delta) + \omega_6 (t|xx),
\end{equation}
where $\omega_i$ are user defined weights. The simplex optimizer will only yield a local minima within a small neighbourhood of the starting potentials. As such, many starting potentials are generated randomly with a sobol sequence \cite{bratley2003implementing} and tested. The metric $\chi$ is calculated for each set of potentials and the 100 simulations with the smallest $\chi$ are selected for further optimization by the simplex optimizer. The resolution is then tested for the top performing candidate. 

\section{Simulation Results}
\label{sec:4}
\begin{table}
\centering
\caption{The optimized and simulated MR TOF configuration for the LPT/MR TOF combination. The mirror potentials have been optimized for isochronous reflection. There may exist better configurations that have not yet been found.}
\label{tab:1}       
\begin{tabular}{ccccccc|cccccc}
\hline
\multicolumn{7}{c |}{Mirror Dimension (mm)} & \multicolumn{6}{| c}{Mirror Potential (V)} \\
\hline
\hline
L1 & L2 & L3 & L4 & L5 & L6 & Radius & V1 & V2 & V3 & V4 & V5 & V6 \\
\hline
15 & 15 & 15 & 25 & 25 & 41 & 18 & 1705 & 1206 & 1072 & 688 & -479 & -2018 \\
\hline
\end{tabular}
\end{table}

\begin{figure}
    \centering
    \includegraphics[width = \textwidth, height = 0.8 \textwidth]{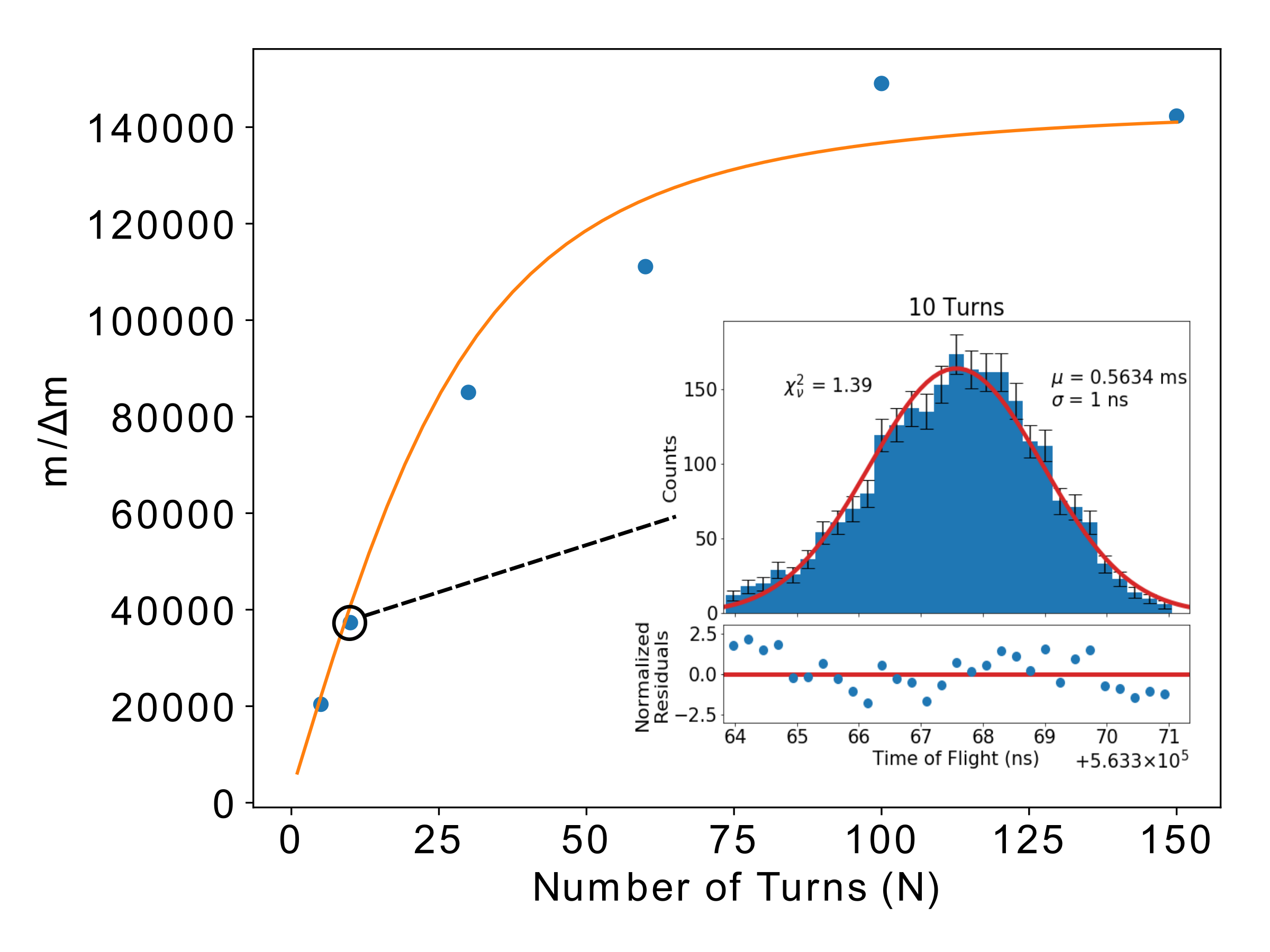}
    \caption{The simulated mass-resolving power as a function of $N$. The TOF spectrum is simulated with 1000 ions for various $N$, where a Gaussian fit is used to calculate the mass-resolving power with $t/(2\Delta t)$ at FWHM. The MR TOF reaches a maximum mass-resolving power of $\approx$ 140,000, with $\Delta T = 0.2$ ps (obtained through fit of eq.  (\ref{resolvingpower})), using the parameters $t_0 = 30$ $\mu$s and $\Delta t_0 = 7$ ns.}
    \label{fig:resplot1}
\end{figure}

\begin{figure}
    \centering
    \includegraphics[width = \textwidth, height = 0.7 \textwidth]{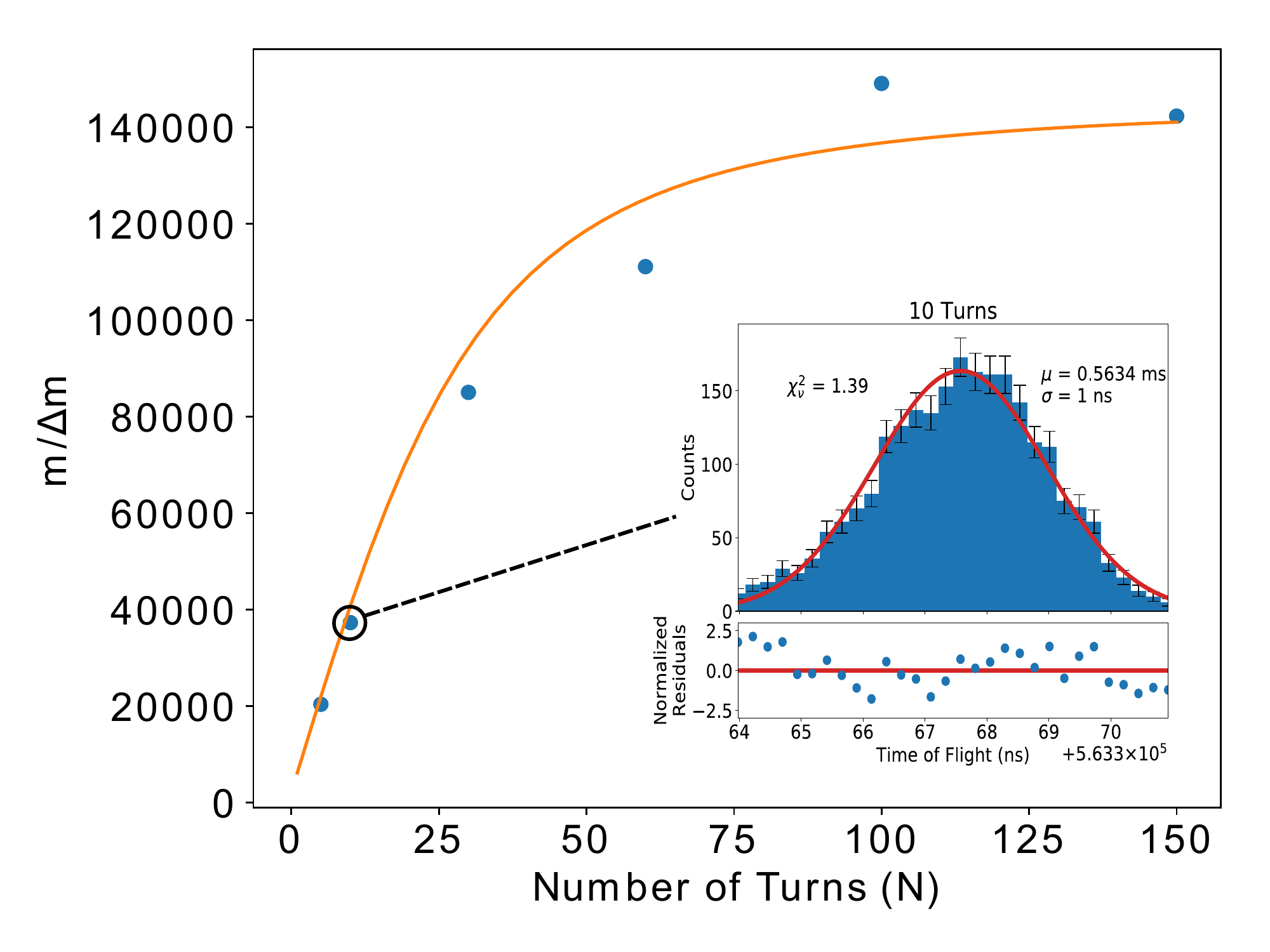}
    \caption{The simulated mass-resolving power as a function of $N$. For $m/\Delta m\approx$140,000, the mass range is $(m/q)_{\text{max}}$/$(m/q)_{\text{min}} \approx 1.01$.}
    \label{fig:resplot2}
\end{figure}

The optimized parameters derived in this work are summarized in Table \ref{tab:1}. The corresponding resolution was calculated by simulating the TOF spectra in SIMION for a varying $N$. A Gaussian function was fit to each of the spectra to extract the mean and FWHM. The calculated mass-resolving power versus $N$ was then fit with eq. (\ref{resolvingpower}), from which, the TOF dispersion introduced with each turn, $\Delta T$, is extracted. As shown in Fig. \ref{fig:resplot1}, the MR TOF can reach a maximum mass-resolving power of $ m/\Delta m\approx$140,000, with $\Delta T = 0.2$ ps, using the parameters $t_0 = 30$ $\mu$s and $\Delta t_0 = 7$ ns. Due to the large computing and memory requirements, the MR TOF geometry was only adjusted with mm precision, which is well above machining tolerances. Geometric parameters were also adjusted independently of each other, and as such, only a small fraction of the parameter space has been explored. It is likely that configurations with better resolving power exist. The mirror potential optimization procedure is currently limited by the $(t|xx)$ parameter that degrades the time-focus if the beam width is over $2$ mm.  \newline

The unambiguous mass range for the MR TOF has been calculated using eq. (\ref{eq:mass_acc}) with $\lambda_{\text{inj}}$=0.63 and $\lambda_{\text{mir}}$ = 0.21, and it is shown in Fig. \ref{fig:resplot2} as a function of $N$. The mass range drops sharply as a function of $N$ and is well approximated by $(N+1)/N$. A range of $(m/q)_{\text{max}}$/$(m/q)_{\text{min}} \approx 1.15$ with $m/\Delta m \approx 50000$ is possible. At $m/\Delta m \approx$140,000, the mass range is $(m/q)_{\text{max}}$/$(m/q)_{\text{min}} \approx 1.01$.

\section{Conclusions}
\label{sec:5}
In broadband mode, the MR TOF can have a mass range of $(m/q)_{\text{max}}$/$(m/q)_{\text{min}} = 1.45-1.15$ with a mass-resolving power of 20,000-50,000, respectively. This resolving power is sufficient to separate Xe from most ions produced by the aforementioned ion sources. In high resolution mode, the MR TOF can have a mass resolving power of 140,000, which is sufficient for isobaric separation, i.e., $^{136}$Xe from $^{136}$Ba. Since the mirror potentials are independent of $N$, the MR TOF has a quickly adjustable mass resolving power and mass range. This performance is sufficient for broadband mass measurements of the proposed ion-extraction process in Ba-tagging. The MR TOF can achieve third order time focusing, $(t|\delta \delta \delta) = 0$, but the resolution is limited by the $(t|xx)$ and $(t|\delta \delta \delta \delta)$ parameters. This issue will be investigated in future work. To mitigate the effect of $(t|xx)$, the ion beam will be collimated to a maximum diameter of 2 mm, which will lower the MR TOF's acceptance of LPT ions to $\approx$ 60$\%$.\newline

\textbf{Acknowledgements}   We thank Timo Dickel, at GSI, and T. Eronen, at the University of Jyv\"{a}skyl\"{a}, for helpful discussions concerning the design of the MRTOF. We also thank Maxime Brodeur and Brad Schulz for supplying information and designs of the MR TOF at the University of Notre Dame. We would also like to thank the members of the nEXO collaboration, for their support and guidance. A special thanks also to Thanh Nguyen, for useful feedback on this manuscript. \newline 

\textbf{Funding}   This work has been supported by NSERC, the Canadian Foundation for Innovation, and the McDonald Institute (CFREF) in Canada. nEXO is supported by the Offices of Nuclear and High Energy Physics within DOE’s Office of Science, and NSF in the United States; by NSERC, CFI, FRQNT, NRC, and the McDonald Institute (CFREF) in Canada; by SNF in Switzerland; by IBS in Korea; by RFBR in Russia; and by CAS and ISTCP in China.
\bibliographystyle{spphys}       
\bibliography{bibl.bib}   

\end{document}